\documentclass[a4paper]{jpconf}
\usepackage{graphicx}

\begin{document}

\def\bea{\begin{eqnarray}}
\def\eea{\end{eqnarray}}
\def\vt{\vartheta}
\newcommand{\rhat}{\hat{r}}
\newcommand{\iotahat}{\hat{\iota}}
\newcommand{\phihat}{\hat{\phi}}
\newcommand{\h}{\mathfrak{h}}
\newcommand{\be}{\begin{equation}}
\newcommand{\ee}{\end{equation}}
\newcommand{\ber}{\begin{eqnarray}}
\newcommand{\eer}{\end{eqnarray}}
\newcommand{\fmerg}{f_{\rm merg}}
\newcommand{\fcut}{f_{\rm cut}}
\newcommand{\fring}{f_{\rm ring}}
\newcommand{\cA}{\mathcal{A}}
\newcommand{\ie}{i.e.}
\newcommand{\df}{{\mathrm{d}f}}
\newcommand{\dt}{{\mathrm{d}t}}
\newcommand{\pj}{\partial_j}
\newcommand{\pk}{\partial_k}
\newcommand{\psifl}{\Psi(f; {\bm \lambda})}
\newcommand{\hp}{h_+(t)}
\newcommand{\hc}{h_\times(t)}
\newcommand{\Fp}{F_+}
\newcommand{\Fc}{F_\times}
\newcommand{\Ylm}{Y_{\ell m}^{-2}}
\def\no{\nonumber \\ & \quad}
\def\noQ{\nonumber \\}
\newcommand{\mc}{M_c}
\newcommand{\vek}[1]{\boldsymbol{#1}}
\newcommand{\vdag}{(v)^\dagger}
\newcommand{\bvtheta}{{\bm \vartheta}}
\newcommand{\btheta}{{\bm \theta}}
\newcommand{\brho}{{\bm \rho}}
\newcommand{\pa}{\partial_a}
\newcommand{\pb}{\partial_b}
\newcommand{\Psieff}{\Psi_{\rm eff}}
\newcommand{\Aeff}{A_{\rm eff}}
\newcommand{\deff}{d_{\rm eff}}
\newcommand{\corr}{\mathcal{C}}
\newcommand{\bvthat}{\hat{\mbox{\boldmath $\vt$}}}
\newcommand{\bvt}{\mbox{\boldmath $\vt$}}

\title{Tackling excess noise from bilinear and nonlinear couplings in  gravitational-wave interferometers}

\author{Sukanta Bose$^{1,2}$, Bernard Hall$^1$, Nairwita Mazumder$^1$,
  Sanjeev Dhurandhar$^2$, Anuradha Gupta$^2$, 
  Andrew Lundgren$^3$}

\address{$^1$ Department of Physics \& Astronomy, Washington State University,
1245 Webster, Pullman, WA 99164-2814, U.S.A}
\address{$^2$ Inter-University Centre for Astronomy and Astrophysics, Post
  Bag 4, Ganeshkhind, Pune 411 007, India}
\address{$^3$ Max-Planck-Institut f\"{u}r Gravitationsphysik (Albert-Einstein-Institut), Callinstrasse 38, 30167 Hannover, Germany} 

\ead{sukanta@wsu.edu}

\begin{abstract}

We describe a tool we improved to detect excess noise in
the gravitational wave (GW) channel arising from its bilinear or
nonlinear coupling with fluctuations of various components of a 
GW interferometer and its environment. We also
describe a higher-order statistics tool we developed to characterize
these couplings, e.g., by unraveling the frequencies of the fluctuations
contributing to such noise, and demonstrate its utility by applying
it to understand nonlinear couplings in Advanced LIGO engineering
data. Once such noise is detected, it is highly desirable to remove it or correct
for it. Such action in the past has been shown to improve the
sensitivity of the instrument in searches of astrophysical signals. 
If this is not possible, then steps must be taken to mitigate
its influence, e.g., by characterizing its effect on astrophysical
searches. We illustrate this through a study of the effect of
transient sine-Gaussian noise artifacts on a compact binary coalescence template bank.

\end{abstract}

\section{Introduction}

The advanced detector era was ushered in by the two Advanced
LIGO (aLIGO) detectors~\cite{TheLIGOScientific:2014jea}  in September 2016. 
With plans afoot for the Advanced Virgo (AdV) 
detector~\cite{TheVirgo:2014hva} to join it this decade, we
expect the era of gravitational wave (GW) astronomy to begin in earnest~\cite{Aasi:2013wya}.
The sensitivity of an interferometer in the detection
band (which in aLIGO is currently between a few tens of Hertz to a few kiloHertz)~\cite{TheLIGOScientific:2014jea}, however, can be affected adversely due to
unwanted noise arising from various sources. The effect of noise
transients (see, e.g., Ref.~\cite{Powell:2015ona} and references
therein) on astrophysical searches was studied in Ref.~\cite{Colaboration:2011np}
for binary neutron signals, in Refs.~\cite{Aasi:2014bqj,Talukder:2013ioa}
for intermediate mass black hole signals, in Ref.~\cite{Aasi:2012rja}
for binary black hole signals, and in Ref.~\cite{Aartsen:2014mfp}
for unmodeled transient signals.
Detector characterization efforts that have been invested toward improving the quality
of GW data in the recent past are described in
Refs.~\cite{Aasi:2014mqd,Nuttall:2015dqa,Aasi:2012wd}.
In this work we improve upon a subset of those efforts,
specifically, in regards to identifying the presence of certain types of
noise that appear in the detection band owing to bilinear or nonlinear
coupling of fluctuations at lower frequencies. We also describe a
higher-order statistics tool we developed to characterize
these couplings by finding the frequencies of the fluctuations that
contribute to such noise. We demonstrate its utility by applying it to
aLIGO engineering data. 

One of the simplest methods used for identifying the source of
excess noise in the GW strain data of a detector is to measure its
coherence with data recorded by, say, a sensor monitoring an
environmental disturbance or a fluctuation 
in an interferometer component. Sensors record 
such data as time-series in multiple channels. For instance, time variations in 
the pitch, yaw, and roll (angular) degrees of freedom of each test
mass can be extracted from the time-series data acquired by the
Optical Sensors and Electromagnetic Actuators (OSEMs) attached
to the multiple stages and levels of their suspension and recorded in
various channels. The power-line harmonics can be 
extracted from the data recorded by the magnetometers. 
On the other hand, an example of an environmental disturbance is
ground motion, which is measured by seismometers and accelerometers.

The coherence statistic is useful in identifying noise sources,
be they internal to the interferometer or environmental, that couple linearly
with the GW channel.
There are various tools available to find them effectively and, therefore, will
not be studied here (see, e.g., Ref.~\cite{Aasi:2014jkh} and
references therein).
Unwanted noise can also arise in the GW channel owing to
higher order couplings among multiple channels. 
For instance, when a servo-control system drifts near the edge of 
its actuation range, a further small fluctuation can drive the 
actuator into a nonlinear regime, causing elevated rate of noise transients
in the GW channel. This can involve multiple channels in a nonlinear fashion,
such as when a large alignment drift reduces the actuation range for 
additional length fluctuations. Another example would be a bilinear 
combination of a slow angular drift in alignment that causes a 
measured light beam to drift off its detector, thus amplifying the 
effect of any further higher-frequency jitter of the light-beam 
alignment. This type of behavior increases the noise background of 
searches for short-duration astrophysical signals, e.g., from binary 
black hole coalescences, hence, reducing the detector's sensitivity to 
them~\cite{Aasi:2014mqd}.

\begin{figure}[h]
\includegraphics[width=20pc]{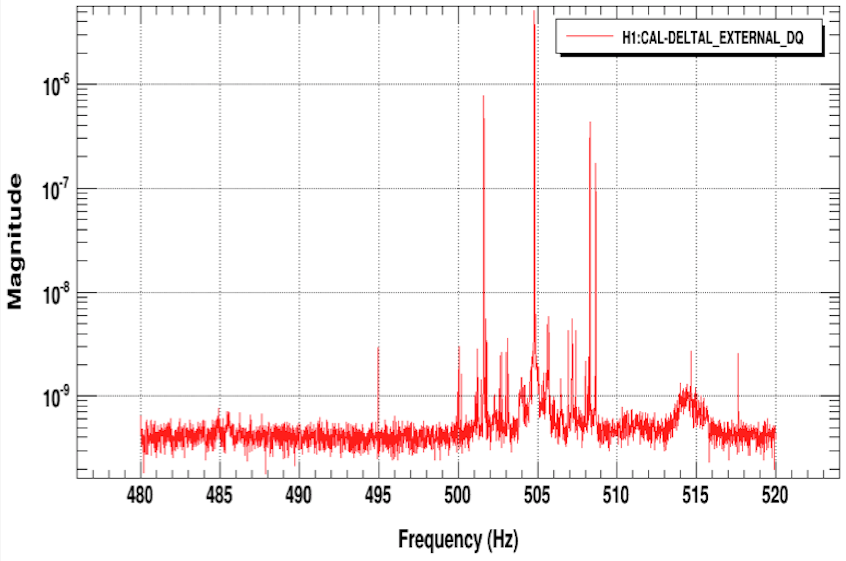}
\caption{Excess noise in the form of side-bands of a violin mode, at
  approximately $505$~Hz, 
  suggests the possible existence of nonlinear coupling between the
  strain data in the GW channel (which is represented here by
  ``H1:CAL-DELTAL\_EXTERNAL\_DQ'') and noise in a host of other
  channels. (The discussion around Eq.~(\ref{eq:sidebands}) explains
  the origin of such side-bands.)}
\label{fig:violinMode}
\end{figure}

The Bilinear Coupling Veto (BCV) tool~\cite{bcv2014} has been shown to
be useful in identifying the presence of non-Gaussian, bilinear component of the
noise to a certain degree.
While there are many tools that can identify the presence of component
frequencies in a given time series, those typically employed are not
able to provide any definitive evidence of nonlinear couplings that
may be present (see, however, Ref.~\cite{Tiwari:2015ofa}).  
For example, the noise power spectral
density (PSD) of the GW channel can show frequencies at which excess
noise due to such couplings is present (see, e.g., the side-bands of
the violin mode of a LIGO test mass in Fig.~\ref{fig:violinMode}). However,
there is no phase information present here, and
one 
needs to employ higher-order statistics, in general, for diagnosing the 
presence of non-Gaussianity and nonlinear couplings~\cite{mendel1991}.
BCV addresses this problem, partially, by finding strong coherence
between the GW time-series, on the one hand, and the product of the
time-series of two interferometer channels, on the other hand. 
Reference \cite{bcv2014} termed this product as a {\em pseudo} 
channel. (Note that the pseudo channel can be constructed from 
environmental channels as well.)
Typically, one of the time-series in that product is taken
from a channel that records slow variation, e.g., the changing
pitch-angle of an end-test mass, and the other time-series is taken
from one that records fast variation, e.g., the signal controlling the
length of the power-recycling cavity. We call this coherence the BCV
statistic. In this work, we use the higher-order statistics tool to
unearth some characteristics of the noise couplings that BCV finds.

\begin{figure}[h]
\includegraphics[width=20pc]{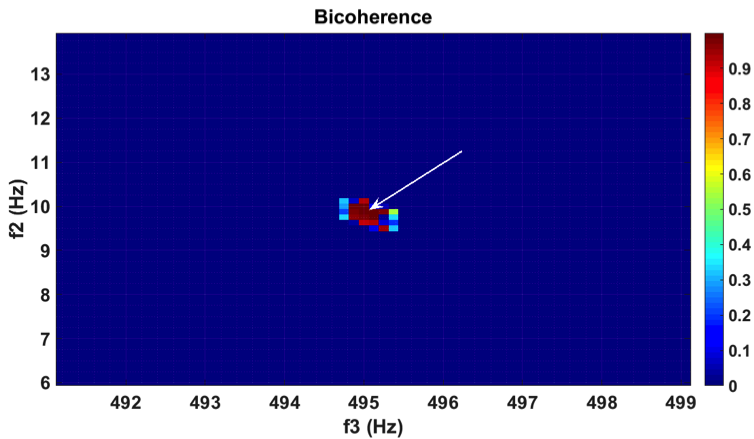}
\includegraphics[width=20pc]{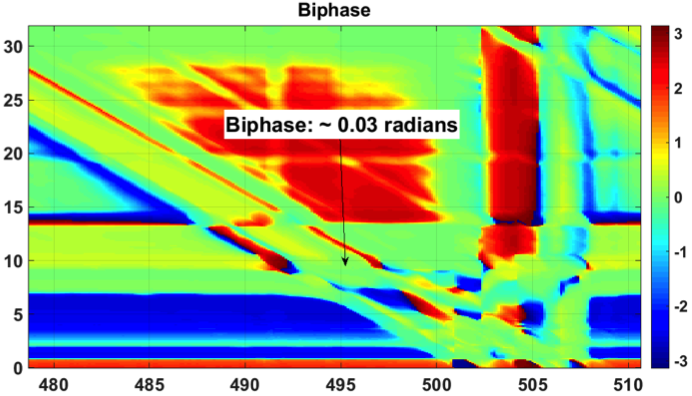}
\caption{The bounce mode ($f_2 \approx$10~Hz) of one of the test masses
  couples with its violin mode ($f_1 \approx$505~Hz) to create a side-band at
  ($f_3 \approx$495~Hz). The color-bar shows the bicoherence and the biphase
  (in radians) in the left and right figures, respectively, computed
  on the GW strain data.
}
\label{fig:bounce}
\end{figure}

\begin{figure}[h]
\includegraphics[width=20pc]{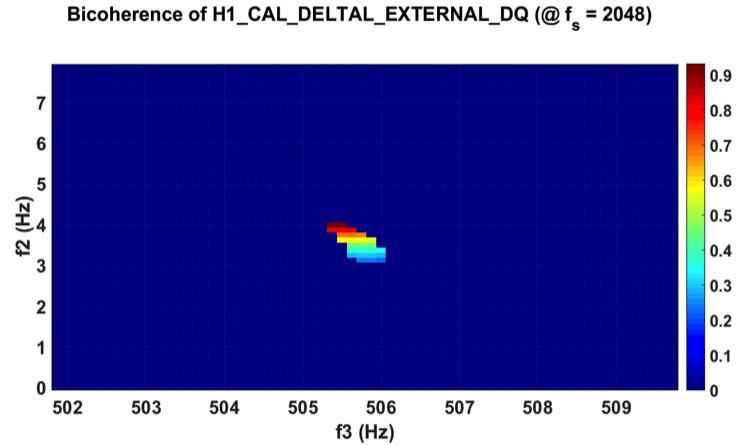}
\includegraphics[width=20pc]{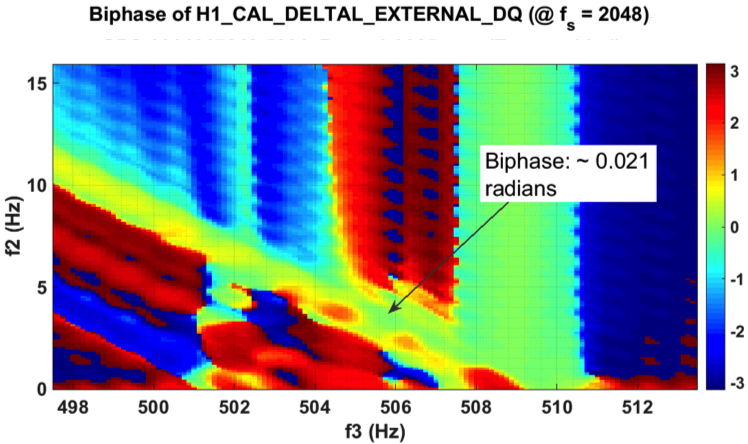}
\caption{A 
mechanical resonance at ($f_2 \approx$3.8~Hz) of the suspension of one of the
  test masses couples with its violin mode ($f_1 \approx$509~Hz) to create a side-band at ($f_3 \approx$505.2~Hz). The color-bar shows the bicoherence and the biphase
  (in radians) in the left and right figures, respectively, computed
  on the GW strain data.
}
\label{fig:suspension}
\end{figure}

\section{Detecting bilinear and nonlinear noise couplings}

The last published work on BCV~\cite{bcv2014} presented results
obtained from its application to LIGO's sixth Science Run ``S6'' data, 
when the wavelet-based Kleine Welle \cite{Chatterji:2004qg}
excess-power detection tool was used to identify noise transients. In
preparation for aLIGO observations, we have upgraded
BCV to work (a) additionally with transients found by ``Omicron'', 
which uses the $Q$-transform on a sine-Gaussian 
basis~\cite{Chatterjithesis:2005,Chatterji:2004qg}, and (b) with aLIGO channels,
which use a different nomenclature and are more numerous than LIGO channels.

We use BCV to identify episodes of bilinear coupling as follows.
First, the occurrence times and other characteristics of excess noise
transients or ``glitchiness'' are identified by Omicron or Kleine
Welle in the GW channel and multiple instrumental and environmental 
channels that record fast variations.
Second, time-coincidences of transients in that latter set of channels 
and the GW channel are
identified for possible correlations between them. Third, during times
of coincidences the BCV statistic is computed on the GW
time-series and a pseudo channel, which is a product of the
time-series with fast variations and a time-series picked from
channels recording slow variations. For every coincidence, the last
step is repeated over multiple channels that record slow variations.
A strong value of the BCV statistic indicates
the presence of bilinear coupling. To assess how large the value of
this statistic can be in the absence of a coupling, the BCV statistic 
is computed also for acausally large relative time-shifts
between the GW channel and the pseudo channel of interest.
(Reference \cite{bcv2014} used the bilinear coherence
statistic for identifying bilinear coupling, but it is possible to replace
it with the {\em bicoherence} statistic defined below for diagnosing 
the presence of nonlinear couplings.) When the BCV
statistic is found to be significantly larger than the background, the
GW data is flagged for further data quality studies and for setting
vetoes for astrophysical searches.  
However, to search for the sources of these transients, one must
unearth more information about their characteristics than just what
channels are involved in producing such bilinear /
multi-linear or nonlinear couplings. For example, the frequencies of
the disturbances that couple provide valuable information about the source
and pathways of such couplings. Below we describe a few of those
statistics that we use to find such information.

\section{Characterizing noise couplings by employing higher order statistics} 

One possible way of probing the nature of beyond-linear-order 
couplings is to compute a higher order statistic, e.g., the bicoherence, of the
disturbances that we suspect can couple with one another. 
Strong bicoherence between channels can suggest couplings of their respective subsystems. For instance, mechanical resonances of structures (e.g., periscopes) show up at higher frequencies in the GW and auxiliary channels as well as some sensor channels.

Mathematically, a set of $N$ disturbances or ``signals'' $x_k$ is 
said to combine linearly when they
superimpose to form a new signal, $z_{\rm lin}$, as follows:
\be 
z_{\rm lin} (t) = \sum_{k=1}^N x_k(t) \,,
\ee 
In the special case where the $x_k$ are sinusoidal, one gets
\be 
z_{\rm lin} (t) = \sum_{k=1}^N a_k \cos (2\pi f_k t+\phi_k)\,,
\ee 
where $a_k$ and $f_k$ are the amplitudes and frequencies of the linearly combining signals,
and $\phi_k$ are their initial phases, respectively. When $N=2$, we
will identify $x_1(t)$ as $x(t)$ and $x_2(t)$ as $y(t)$, and write $z_{\rm lin} (t) = x(t) + y(t)$.

Though there are different means by which signals may combine
nonlinearly, we will consider one of the simplest yet ubiquitous
coupling, namely, a type of coupling known as the quadratic
phase coupling (QPC)~\cite{fackrell1996}.
The reason for this choice is that it is the most significant term in
many nonlinear couplings and is also straightforward to model.
In this case, two (or more) signals combine to form a new signal, $z(t)$. For a pair of
coupled signals one gets,
\bea\label{eq:sidebands}
z(t) &=& p [ x(t) + y(t)]^2 + q [x(t) + y(t)] \nonumber \\
&=& p\left[ a_1 \cos (2\pi f_1 t+\phi_1) + a_2 \cos (2\pi f_2 t+\phi_2)
\right]^2 \nonumber \\
 &&+ q \left[ a_1 \cos (2\pi f_1 t+\phi_1) + a_2 \cos (2\pi f_2 t+\phi_2)
\right] 
\,,
\eea
where $p$ and $q$ are real numbers, and the pair of frequencies, $(f_1, f_2)$, is known as a
bifrequency. Heretofore, $t$ will denote a time index. Since the
right-hand side has terms that denote oscillations (and power) at frequencies
$(f_1\pm f_2)$, $2f_1$, and $2f_2$, in order to detect QPC it is necessary to
use a statistic that can aid the detection of these frequency
components. In some of the examples studied in this paper, $f_2$ is
much smaller than $f_1$. In such an event, excess noise will appear in
the sidebands $(f_1\pm f_2)$ of $f_1$.

To obtain one such statistic, we first divide the time-series, $x(t)$,
$y(t)$, and $z(t)$ into $M$ segments $x_\alpha(t)$, $y_\alpha(t)$, and
$z_\alpha(t)$, respectively, where $\alpha$ is the segment index. Next
the Fourier transforms, $X_\alpha(f)$, $Y_\alpha(f)$, $Z_\alpha(f)$, 
of each of these segments is taken, respectively, followed by the computation of the average of the quantity
$X_\alpha(f_1) Y_\alpha(f_2) Z_\alpha^*(f_1 + f_2)$ over those
segments:
\be
B(f_1, f_2) = \frac{1}{M}\sum_{\alpha=1}^M X_\alpha(f_1) Y_\alpha(f_2) Z_\alpha^*(f_1 +
f_2)\,.
\ee
A normalized and real function derived from $B(f_1, f_2)$ is defined as 
\be 
b(f_1, f_2) =\left| B(f_1, f_2) \right| \left[\frac{1}{M} \left|
    \sum_{\alpha=1}^M X_\alpha(f_1) Y_\alpha(f_2) \right|  \frac{1}{M} \left|
    \sum_{\alpha=1}^M Z_\alpha^*(f_1 +f_2)\right|  \right]^{-1}\,,
\ee 
which takes values in the range $[0,1]$. It attains large values when 
$p$ is large, i.e., when a strong QPC is present. 
Another related statistic,
defined for individual segments, is 
\be\label{biphase}
\Delta\phi_\alpha(f_1, f_2)  = \arg\left[X_\alpha(f_1) Y_\alpha(f_2)  Z_\alpha^*(f_1 +f_2)\right]\,.
\ee
Its average,
$\Delta\phi$, for a large enough $M$, should be at the
level of the noise when a QPC is present. Otherwise it can be larger,
depending on the type of oscillations present.
The statistic $B(f_1, f_2)$ is related to 
the bispectrum, $B_Z(f_1, f_2)$, which is obtained by replacing 
$X_\alpha(f_1)$ and $Y_\alpha(f_2)$ in $B(f_1, f_2)$ by $Z_\alpha(f_1)$
and $Z_\alpha(f_2)$, respectively. The same replacements in $b(f_1,
f_2)$ and $\Delta\phi$ yield the bicoherence, $b_Z(f_1, f_2)$ and
biphase $\Delta\phi_Z$.
Thus, for finding QPCs we search for frequency triplets,
$(f_1,f_2,f_3)$ for which the bicoherence is high (i.e., close to
unity) and the biphase is small (i.e., close to noise level).

In Fig.~\ref{fig:bounce} we show the
bicoherence $b_Z(f_1, f_2)$ and the biphase $\Delta\phi_Z$, where $Z$
is the Fourier transform of the GW strain data,  in terms of
$(f_3,f_2)$ for a stretch of time that was flagged by BCV as a
likely candidate for the presence of bilinear or nonlinear coupling
noise. Here the bicoherence is strong at $(f_3,f_2)=(495,10)$~Hz. 
This indicates that excess noise is being created by a QPC-like 
nonlinear coupling between $f_2 \approx 10$~Hz, which is known to be a bounce
mode of a test mass, and $f_1 = f_2+f_3 \approx 505$~Hz, which is the
violin mode of the same test mass.
In Fig. \ref{fig:suspension} we show plots of the same two quantities
on a stretch of data where the bicoherence is strong at
$(f_3,f_2) \approx (505.2,3.8)$~Hz. This is strong evidence for
nonlinear coupling noise between $f_2 \approx 3.8$~Hz, which is known to
be a mechanical resonance 
of a test mass, and $f_1 = f_2+f_3 \approx 509$~Hz, which is the
violin mode of that test mass. (Note that different test mass suspensions have
somewhat different violin mode frequencies.)

\section{Effect of noise artifacts on astrophysical searches}

Once a strong indication of bilinear or nonlinear coupling is found as the
potential source of a noise artifact, steps must be taken to establish
the causal nature of the effect, find its
origin and then either remove it or correct its behavior, as the case
may be. (More details on this procedure will be reported in a future work, where
the improvement in the sensitivity of specific astrophysical searches
in aLIGO data brought about by using BCV will be provided.)
When it is not possible to completely remove it or correct it 
steps must be taken to mitigate its effect on the GW channel or on
astrophysical searches. A brief example of one such step is presented
in this section. 

\begin{figure}[h]
\includegraphics[width=20pc]{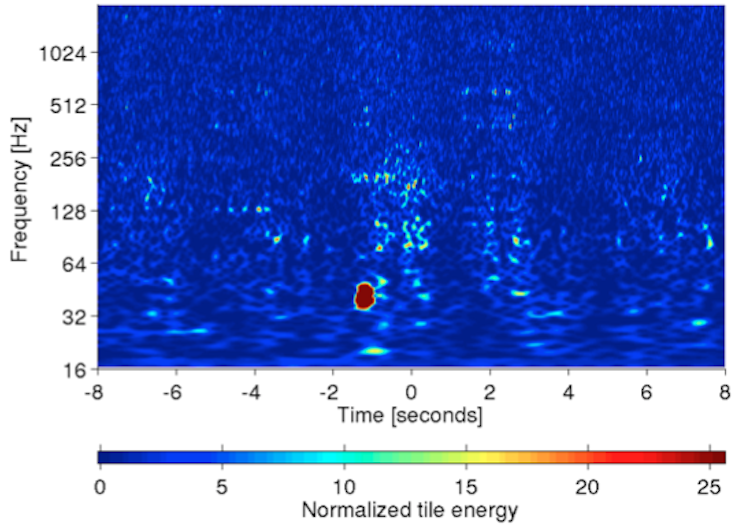}
\includegraphics[width=20pc]{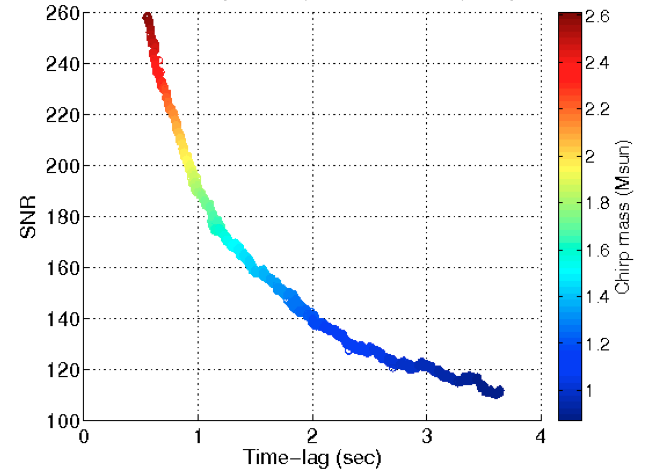}
\caption{A noise glitch found by Omicron (left figure) and the effect of one
such glitch, as modeled by a sine-Gaussian~\cite{tito2014} with
quality factor, $Q=10$ and $f_0=100$~Hz, on a
low-mass CBC template bank (right figure). The time-lag in the latter
figure is the time duration between the central time of the glitch
and the end-time of the CBC template triggered by that glitch. The
signal-to-noise ratio (SNR) of a glitch is maximum when the glitch
overlaps with that part of the template where the instantaneous
frequency of the latter equals $f_0$. As the
chirp-mass decreases the time-lag increases because the duration
of the template after an instantaneous frequency of $f_0$
increases.
}
\label{fig:omicron}
\end{figure}

Increased noise, e.g, in the form of side-bands of the violin modes or line
harmonics, that is reasonably stationary can hurt the sensitivity of a
search for continuous-wave signals from spinning neutron
stars~\cite{Abadie:2011wj} or,
possibly, a stochastic GW background~\cite{Aasi:2014jkh,Talukder:2014eba}.
On the other hand, transient noise artifacts can adversely affect the sensitivity of
astrophysical burst or compact binary coalescence (CBC) 
searches by increasing their background~\cite{Aasi:2014mqd,Aasi:2014tra,Dayanga:2013tta}.
For CBC searches, signal discriminatory tests, such as
the chi-square test~\cite{Allen:2004gu}, help in keeping the background under control,
especially for low-mass searches. For high-mass searches, where the
search templates~\cite{Aasi:2012rja} are short in duration and have very few cycles, much
like some of the noise transients themselves,
these tests are known to be less powerful, and other tests must be devised to
supplement them. One such idea was explored in
Ref.~\cite{tito2014}, which found that for a class of these noise
artifacts that can be modelled as sine-Gaussian bursts, the time-lag of the CBC
templates that get triggered by a burst depends in a
(semi-analytically) predictable way on the burst's central frequency
and the template's chirp mass. (The chirp mass of a binary with
component masses $m_1$ and $m_2$ is defined
as $(m_1m_2)^{3/5}/(m_1+m_2)^{1/5}$.) As shown in Fig.~\ref{fig:omicron}, the
SNR of a trigger from such a glitch falls off with
decreasing template chirp-mass (and increasing time-lag). 
For a CBC signal, the SNR typically falls much faster with time-lag
(see, e.g., Ref.~\cite{Sathyaprakash:1991mt}). These characteristics
can be capitalized upon by machine-learning algorithms to complement the
ability of traditional chi-square tests to discriminate such glitches
from real CBC signals \cite{privateDent,boseEtAlCSG}.

\section{Conclusion}

Coupling of disturbances at low frequencies,
such as the bounce and roll modes, and those at higher frequencies,
e.g., the violin modes or power-line harmonics, can adversely affect the detector 
sensitivity in the band where we expect to find astrophysical signals.
Noise of this type can create a stable background 
or one that ebbs and flows slowly with changing anthropogenic,
environmental (e.g., wind-speed variation, microseismic,
etc.) or operating conditions, unless it is recognized and 
removed. Here we demonstrated how BCV can find the presence of such
noise couplings and showed how higher order statistics can be used to
diagnose certain aspects of those couplings. By way of their power for
finding bilinear and nonlinear couplings and diagnosing some of 
their properties these statistics make useful additions to the set of 
detector characterization tools that are already being employed for 
improving aLIGO and AdV data quality. Characterization of
interferometer subsystems and environmental disturbances, and
reduction of low-level correlated noise, are critical to ensuring good sensitivity to
astrophysical searches. With the advent of the first observation run
in the advanced detector era, its need has never been more important.

\ack

We would like to thank Fred Raab for extensive discussions and
guidance. We thank T. Isogai, N. Christensen, P. Ajith for
collaboration on the BCV code. We also thank T. Dal Canton, P. Fritschel, D. Macleod,
T. J. Massinger, L. Nuttal, S. Penn, P. Saulson, R. Schofield, D. Sigg, J. Smith, and
S. Whitcomb for helpful discussions. Finally, we thank Jess McIver for
discussions and for carefully reading the manuscript and making
several useful and extensive comments on it. This work was supported in part
by NSF awards PHY-1206108 and PHY-1506497. The LIGO Document ID of
this paper is LIGO-P1500181.

\end{document}